# Strong Coupling of Self-Trapped Excitons to Acoustic Phonons in Bismuth Perovskite $Cs_3Bi_2I_9$

Xing He,[a] Naveen Kumar Tailor,[b] Soumitra Satapathi,[b] Jakoah Brgoch,[c] Ding-Shyue Yang*[a]

[a] *Department of Chemistry, University of Houston, Houston, Texas 77204, United States of America*

[b] *Department of Physics, Indian Institute of Technology Roorkee, Roorkee, Uttarakhand 247667, India*

[c] *TcSUH and Department of Chemistry, University of Houston, Houston, Texas 77204, United States of America*

*To whom correspondence should be addressed. Email: yang@uh.edu

**Abstract**

To assess the potential optoelectronic applications of metal-halide perovskites, it is critical to have a detailed understanding of the nature, strength, and dynamics of the interactions between carriers and the polar lattices. Here, we report the electronic and structural dynamics of bismuth-based perovskite $Cs_3Bi_2I_9$ revealed by transient reflectivity and ultrafast electron diffraction. A cross-examination of these experimental results combined with theoretical analyses allows the identification of the major carrier–phonon coupling mechanism and the associated time scales. It is found that carriers photoinjected into $Cs_3Bi_2I_9$ form self-trapped excitons on an ultrafast time scale. However, they retain most of their energy and their coupling to Fröhlich-type optical phonons is limited at early times. Instead, the long-lived excitons exert an electronic stress via deformation potential and develop a prominent, sustaining strain field as coherent acoustic phonons in 10 ps. From sub-ps to ns and beyond, a similar extent of the atomic displacements is found throughout the different stages of structural distortions, from limited local modulations to a coherent strain field to the Debye–Waller random atomic motions on longer times. The current results suggest the potential use of bismuth-based perovskites for applications other than photovoltaics to take advantage of carriers' stronger self-trapping and long lifetime.

**Introduction**

Research activities on metal-halide perovskites (MHPs) have escalated exponentially over the past decade, with tremendous attention especially on their optoelectronic performances but also on other opportunities such as energy storage, radiation detection and photocatalysis.[1] Fundamentally, wide-ranging compositions, crystal structures, electronic properties, and carrier mobilities and lifetimes with different dimensionalities are seen among the great number of MHPs that have been synthesized.[1d, 1e] At the heart of their optoelectronic phenomena, dynamical behaviors including ultrafast carrier dynamics,[2] carrier–phonon couplings,[3] motions of organic cations,[4] and lattice dynamics[5] require many detailed studies for a coherent understanding. Time-resolved pump-probe techniques are particularly crucial to further reveal the interplays among the many essential factors. For example, a number of ultrafast studies took a closer look at the nature of the interactions between photoinjected carriers and the polar lattice of a MHP bulk to understand charge relaxation and transport. The formation of Fröhlich-type large polarons has been recognized as a leading mechanism for long-lived carriers in lead-halide perovskites (LHPs).[6] In single-crystalline $CH_3NH_3PbI_3$, Mante *et al.* discussed about the coupling between photoinjected charges and acoustic phonons, although the amplitude of the coherent acoustic phonon (CAP) signal is two orders-of-magnitude lower than that of the initial electronic response from transient reflectivity.[7] Very recently, visualization of the dynamic lattice distortions on the nanoscale, including excitation-induced polarons and later-time CAPs, was achieved by using ultrafast x-ray diffuse scattering and diffraction measurements on $CH_3NH_3PbBr_3$.[5] These studies and many others have offered illuminating insights about LHPs, which are frequently used in the comparisons with the properties and behavior of newly developed MHPs.

The toxicity and stability of lead-containing compounds and strong regulatory requirements for their use in consumer products have prompted the search for environmentally-friendly lead-free MHPs with comparable optoelectronic performances.[8] Bismuth-based MHPs received notable attention because they can address the aforementioned concerns. However, the strong interactions between carriers and the polar lattice of Bi-based semiconductors appear to fundamentally limit their optoelectronic performances.[9] Specifically, Wu *et al.* reported experimental evidence from time-resolved optical spectroscopy measurements to infer strong carrier self-trapping in $Cs_2AgBiBr_6$ via deformation potential coupling with acoustic phonons, thus casting doubts on the potential photovoltaic use of Bi-based double perovskites with a soft lattice.[10] However, no diffraction-based direct probing of the structural dynamics has been reported for Bi-based MHPs to date.

Here, using transient reflectivity (TR) and ultrafast electron diffraction (UED), we present the experimental results for the electronic and structural responses of $Cs_3Bi_2I_9$, a quasi-zero-dimensional (quasi-0D) MHP with separated bioctahedral $[Bi_2I_9]^{3-}$ units, following the above-gap injection of photocarriers. A cross-examination of both the optical and diffraction data proves to be critical to revealing the nature of carrier–lattice interactions on ultrashort time scales. The immediate exciton self-trapping in $Cs_3Bi_2I_9$ without much local structural distortion in the first 10 ps indicates a limited coupling to the optical phonons and the retaining of most energy by the trapped carriers without extensive polaronic relaxation. However, the prominent CAPs observed by TR signify a strong coupling of carriers with the acoustic phonons via deformation potential in the first 10 ps, resulting in a strain field persistent beyond the experimental observation window. Additionally, the injected photocarriers at an 8% excitation level are surprisingly long-lived, whose annihilation and nonradiative energy transfer causes the increase of the lattice temperature and therefore random atomic motions on sub-ns to ns time.

**Results and Discussion**

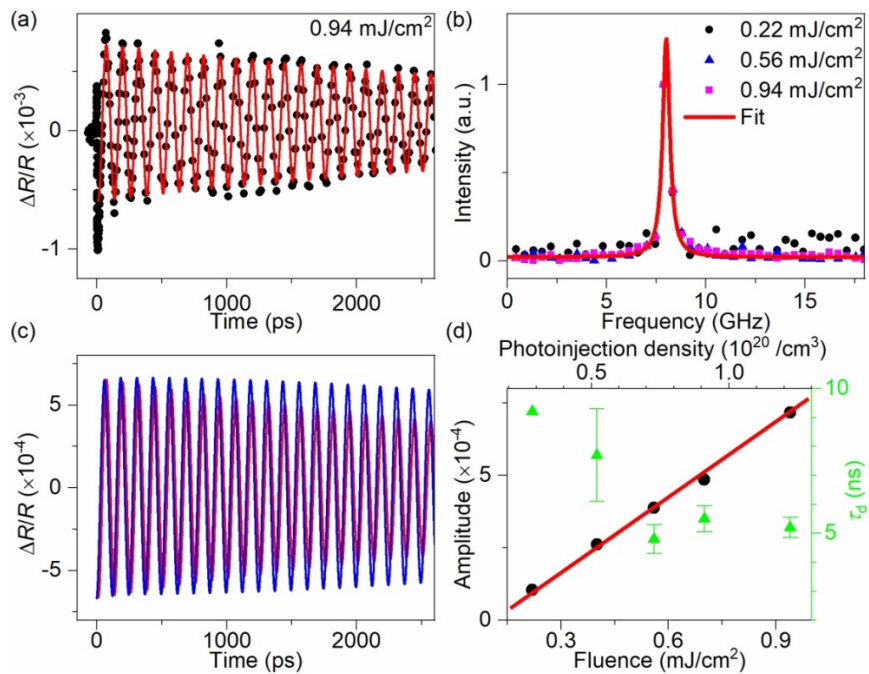

**Figure 1.** Transient reflectivity measured from $Cs_3Bi_2I_9$(001). (a) Prominent, persisting periodic modulations due to carrier-induced CAPs. The red curve is a fit using Equation 1. (b) Fourier-transformed spectrum of the oscillatory signals acquired at three different fluences. The red curve is a Lorentzian fit centered at the frequency matching the period in (a). (c) Comparison of the experimental (red) and theoretical (blue) CAP results. The ~10-ps shift between the two curves signifies the CAP coupling and development time. (d) Fluence dependence of the amplitude (black dots) and decay time (green triangles) of the CAP periodic modulations. The red line is a linear fit intercepting at the origin.

Shown in Figure 1 are the TR results measured in ultrahigh vacuum from freshly cleaved (001) surfaces of crystalline $Cs_3Bi_2I_9$, where the injection of photocarriers is made by using 515-nm light. Given that the energy dispersion of the electronic bands of $Cs_3Bi_2I_9$ is small[11] and the photon energy of $E_{ex}$ = 2.41 eV matches with the vertical transitions across the gap reasonably well, the above-gap photoexcitation results in an initial injection of carriers over a broad range in the Brillouin zone (Figure S1). The wavelength of the probe beam is centered at $\lambda_{pr}$ = 1030 nm, whose photon energy of 1.2 eV is appreciably lower than the indirect band gap. Hence, according to the developed theory for carriers in a band picture,[12] the anticipated optical pump–probe TR observation would be prominent early-time changes with electronic origins followed by relaxation dynamics and (often smaller) lattice contributions resulting from carrier–phonon coupling, which have been seen and analyzed for conventional photovoltaic semiconductors such as silicon, gallium arsenide, and cadmium chalcogenides.[13] However, instead, the TR dynamics of $Cs_3Bi_2I_9$ show an initial change limited in both magnitude and duration followed by prominent and persistent periodic modulations over the measured temporal window (Figure 1a). In what follows, together with the UED results, we will show that these are the signatures for a strong coupling of carriers to acoustic phonons in $Cs_3Bi_2I_9$.

At the excitation fluence of 0.94 mJ/cm², the clear oscillation in the TR trace ($\Delta R/R \sim 10^{-3}$) is due to the interference between the reflected probe beams from the crystal surface and a strain pulse propagating into the bulk at the longitudinal speed of sound $v_{s,\parallel}$.[14] Such laser-induced strain pulses are termed coherent acoustic phonons (CAPs), which arise from the electronic and thermal stresses owing to, respectively, the deformation potential coupling due to long-lived photocarriers and the thermoelastic effect by the excess energy following carrier–phonon coupling and carrier relaxation.[14] A fit of the periodic modulations to the model

$$\frac{\Delta R(t)}{R_0} = A_{osc} \cos\left(\frac{2\pi t}{\tau_{osc}} - \phi\right) \exp\left(-\frac{t}{\tau_d}\right) \qquad (1)$$

yields the initial amplitude $A_{osc} = (7.17\pm0.12)\times10^{-4}$, the period $\tau_{osc}$ = 124.9±0.1 ps, the phase $\phi$ = 0.52 (equivalent to an onset time of oscillation $t_{0,osc} \approx 10$ ps), and the dephasing time $\tau_d \approx 5$ ns that exceeds the observed temporal window (Figure 1a). The Fourier-transformed spectrum also shows a single peak at $1/\tau_{osc}$ = 8.006 GHz, which is independent of the fluence used at room temperature (Figure 1b). Hence, based on the experimental condition,

$$\tau_{osc} = \frac{\lambda_{pr}}{2v_{s,\parallel}\sqrt{n_0^2 - \sin^2\theta_{in}}} \tag{2}$$

where the incidence angle of the probe beam is $\theta_{in} = 35°$ and the refractive index $n_0$ at $\lambda_{pr} = 1030$ nm is assumed to be in the range of 2.08[11a] to 2.20,[15] and we obtain $v_{s,\parallel} = 1.94$ to 2.06 km/s in good agreement with previous reports[15-16] and comparable to (or slightly lower than) that of $CH_3NH_3PbI_3$[7] and bismuth-based double perovskite $Cs_2AgBiBr_6$.[10] Additionally, we find that $A_{osc}$ is linearly dependent on the initial photoinjection density, whereas the dephasing time is similar in the fluence range used or slightly increased at a lower photoexcitation level (Figure 1d).

We conduct theoretical calculations to further understand the origin and magnitude of the prominent CAPs. Given the material's band gap of $E_g \cong 1.9$ eV,[17] the ratio between the electronic and thermal stresses is approximately given by $(\partial E_g/\partial p)\rho C/[3\beta(E_{ex} - E_g)]$ with the mass density $\rho = 5.02$ g/cm³, the specific heat $C \approx 0.178$ J/(g·K) as in the high-temperature limit, and the thermal expansion coefficient $\beta = 4.8\times10^{-5}$ K$^{-1}$.[16, 18] However, as will be shown later from the UED results, carriers still retain most of the above-gap excess energy and effectively diminish $E_{ex} - E_g$ in the denominator. This signifies the dominance of the electronic stress if the carriers have a sufficiently long lifetime.[14] Shown in Figure 1c is the comparison of the experimentally observed periodic modulations with the theoretical curve (see Table S1 for the parameters used), which assumes an instantaneous and persisting lattice deformation and involves a finite spectral width of the probe beam.[14b] Two points are worth noting. First, the observed dephasing cannot be accounted for by the limited attenuation resulting from the finite probe bandwidth. Beside the possible minor inhomogeneity inside the crystalline sample, the loss of carriers (and hence the reduction in the electronic stress) and residual below-gap absorption due to the existence of defects (and therefore an attenuation of the probe beam) may also cause the decrease of the reflected signal. We estimate an upper bound of ~500 cm$^{-1}$ at 1030 nm by translating $\tau_d$ into a round-trip effective depth. Second, the offset in $\phi$ (and hence nonzero $t_{0,osc}$) signifies the time of ~10 ps needed for the carriers' coupling to CAPs to mature, which is significantly longer than the typical carrier relaxation time toward band edges and the offset time observed in conventional semiconductors.[14b] Such a result hints the limitation of the use of a band picture for the photodynamics of $Cs_3Bi_2I_9$ (see below).

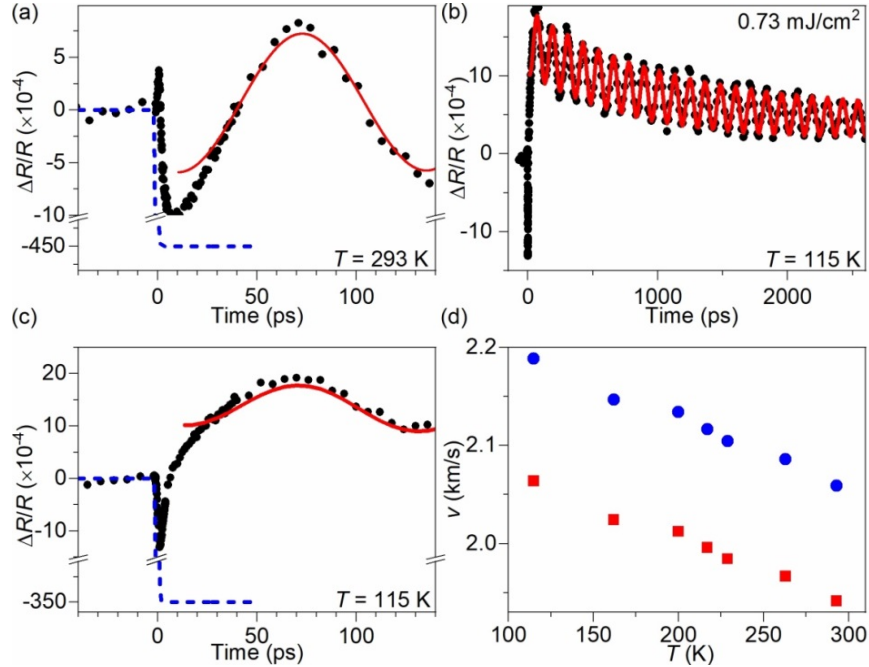

**Figure 2.** Transient reflectivity at early times and select temperatures. (a) Comparison between the experimental results (black dots) and the theoretical free-carrier absorption (blue dashed line) at room temperature. The red curve is the same as that in Figure 1a. (b) Transient reflectivity measured at 115 K. The red curve is a fit using Equation 1. (c) Early-time dynamics of (b), with the theoretical free-carrier absorption shown as the blue dashed line. (d) Temperature dependence of the longitudinal speed of sound obtained from the oscillation period using Equation 2. The blue and red points at a temperature indicate the range calculated using different values for the refractive index.

An analysis of the initial TR signal provides further details about the nature and dynamics of the injected photocarriers. With the *s*-polarized probe,

$$\frac{\Delta R}{R_0} = \text{Re}\left(\frac{4n_0 \cos\theta_{\text{in}}}{(n_0^2 - 1)\sqrt{n_0^2 - \sin^2\theta_{\text{in}}}} \cdot \Delta n\right). \tag{3}$$

For conventional semiconductors with a band picture, three carrier-induced changes to the index of refraction ($\Delta n$) can be modeled for a below-gap probe wavelength: free-carrier absorption (i.e., plasma effect) and band-filling (Bursteirn–Moss effect) give negative contributions, whereas bandgap renormalization leads to a positive change.[12] This means that a negative TR signal is anticipated due to free-carrier absorption within 100s of fs to first ps before photocarriers reach thermalization near the band edges and the contributions from the other two effects become fully matured. At a 515-nm fluence of 0.94 mJ/cm$^2$, we estimate an average photoinjection density of $N_0 \cong 1.2\times10^{20}$ cm$^{-3}$ electron–hole pairs within the optical probed range $\lambda_{\text{pr}}/4\pi|n_0| \cong 40$ nm and hence anticipate a large initial reflectivity drop $\Delta R/R_0 \cong -4.5\%$ from Cs$_3$Bi$_2$I$_9$. However, the opposite sign with a magnitude two orders lower is observed at room temperature (Figure 2a); similar observations at a much lower laser fluence were reported.[17b] In fact, the initial peak and the oscillatory amplitude $A_{\text{osc}}$ are nearly comparable as opposed to those of Cs$_2$AgBiBr$_6$ and CH$_3$NH$_3$PbBr$_3$ with a clearly lower oscillation-to-peak ratio (see Figure S13 of Ref. [10]). At a low temperature $T = 115$ K, the sign of the initial TR change turns out to be consistent with the free-carrier absorption effect, but the magnitude is still ~1.5 orders lower than theoretically predicted (Figure 2, b and c).

Such TR observations signify the dominant presence of bound excitons in Cs$_3$Bi$_2$I$_9$ on the ultrafast temporal scale, whose self-trapped condition effectively removes carriers from the plasma and therefore results in the notable failure of the free-carrier model under a band picture. This is in line with the findings for bismuth double perovskite Cs$_2$AgBiBr$_6$,[10] but we consider stronger trapping in Cs$_3$Bi$_2$I$_9$. According to the Saha equation for the thermal equilibrium established between the photoexcited bound excitons ($n_{\text{EX}}$ as the density) and unbound electron-hole plasma ($n_{e,h}$ as the density),

$$\frac{n_{e,h}^2}{n_{\text{EX}}} = \left(\frac{2\pi\mu_{\text{EX}}k_BT}{h^2}\right)^{3/2} e^{-E_b/k_BT} \qquad (4)$$

where $\mu_{\text{EX}}$ and $E_b$ are the effective reduced mass and the binding energy of the excitons, $k_B$ the Boltzmann constant, and $h$ the Planck constant.[19] For the ranges of photoinjection density and sample temperature used here, $n_{e,h}$ is many orders of magnitude smaller than $n_{\text{EX}}$ given that $E_b$ is close to 300 meV.[20] Thus, the experimental results again support the theoretical model that essentially all photoinjected carriers quickly form bound excitons, which may have its origin in the quantum confinement effect from the quasi-0D bioctahedral [Bi$_2$I$_9$]$^{3-}$ units.[20a] The structural phase transition at 220 K[21] does not appear to cause major effects on the photodynamics. The longitudinal speed of sound derived from $\tau_{\text{osc}}$ at different temperature is shown in Figure 2d, whose dependence agrees well with a previous report.[16]

    Now we turn our focus to the nature of carrier–phonon coupling and the corresponding dynamics. In conventional inorganic semiconductors, the initial coupling is often via optical phonons followed by phonon scattering and thermalization in the acoustic branches. In MHPs, the Fröhlich interaction between carriers and longitudinal optical (LO) phonons has been considered crucial.[6c, 22] We obtain the coupling constants of ~3.0 for electrons and ~3.4 for holes, averaged from the calculated values using different effective masses and dielectric constants reported in the literature (see Table S1 for the parameters used) with the $A_1'$ symmetric stretch of terminal Bi–I bonds.[18b] The results correspond to the large-intermediate Fröhlich coupling similar to or slightly larger than those for CsPbBr$_3$[6c] and Cs$_2$AgBiBr$_6$.[10] Thus, from the viewpoint of LO-phonon coupling, it may not be clear why Cs$_3$Bi$_2$I$_9$ exhibits distinct

dynamics with prominent CAPs and very limited initial electronic TR response. The low mobilities of electrons and holes are consistent with their coupling to the polar lattice and fast self-trapping, although the large effective masses especially along the out-of-plane Γ–A direction also have considerable effects.[23] We note that there is a delay of ~10 ps between the photoinjection time and the time when the TR signal can be largely described by CAPs (Figure 2, a and c). Such a temporal window can be assigned as the duration for the initial carrier–LO phonon coupling; however, we will soon present evidence that LO phonons may not play a dominant role.

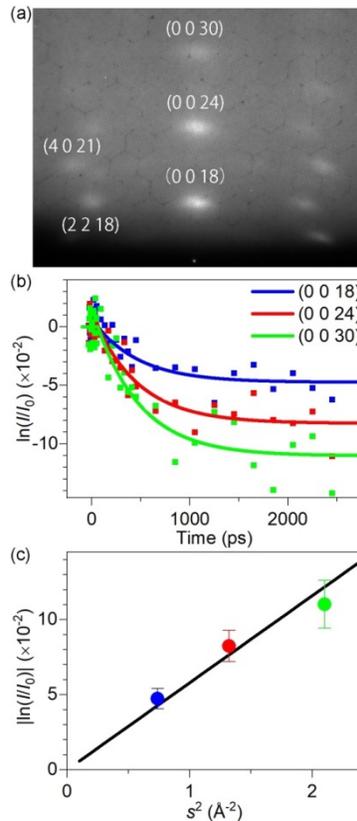

**Figure 3.** UED results measured from $Cs_3Bi_2I_9$(001). (a) Diffraction image recorded at 115 K without photoexcitation. Miller indices are assigned for select Bragg spots. (b) Time-dependent intensity changes of the center three diffraction spots. The solid curves are fits of a single-exponential decay function. (c) Dependence of the intensity decrease on $s_\parallel^2$. The black line is a linear fit intercepting with the origin. The error ranges are ±2 standard deviations obtained from the fits in (b).

We use UED in the reflection geometry to directly probe and reveal the phonon dynamics of $Cs_3Bi_2I_9$ at 115 K. Shown in Figure 3a is the diffraction image with the use of a gently converging beam (to achieve the smallest probe footprint on the sample) without photoinjection, where multiple orders of (006) are seen along the center streak and both (10) and (11) in-plane zones are visible; the in-plane indices follow the room-temperature structure for simplicity. This pattern shows that the freshly cleaved surface is (001) with ordered stacking along the surface normal $c$ axis, whereas the crystal growth along the horizontal directions involves multiple zones and hence crystalline domains of the low-temperature phase. Therefore, we concentrate on the changes of the (00$l$) intensities as a function of delay time, which are directly affected by the out-of-plane components of laser-initiated atomic motions. Rather surprisingly, we do not observe the anticipated fast intensity changes in the first few tens of ps that are very common in UED studies of perovskites[24] and other materials.[25] Instead, the diffraction intensities decrease over a much longer time, whose fits to a single-exponential function yield the long time constant of ~450 ps (Figure 3b). After ~2 ns, the intensity drops reach the full extent and remain so within the observed temporal window. We confirm the structural origin, not any surface transient electric field effect, for the observed diffraction changes, as evidenced by the linear relation between $\ln[I_0/I(t)]$ and $s_\parallel^2$, where $I_0$ and $I(t)$ are, respectively, the intensities of a diffraction spot before the zero of time and at time $t$ after excitation and $s_\parallel = l \cdot c^*$ is the momentum transfer parallel to the surface normal with $c^* = 1/c = 0.0472$ Å$^{-1}$ being the corresponding reciprocal primitive cell length (Figure 3c; see Supplementary Information for the equations and further discussion).[25b, 25d, 26] Additionally, we do not notice clear differences in the time-dependent changes of side Bragg spots such as (2 2 18) and (4 0 21).

The aforementioned time-dependent diffraction changes reveal multiple unique features of the photoinduced structural dynamics of $Cs_3Bi_2I_9$. First, according to the general understanding of carrier–phonon coupling, if the band picture is adequate for a photoexcited material, at least the excess energy above the bandgap ($E_{\text{ex}} - E_g$ per photon) should be quickly transferred to the lattice, which would result in thermal atomic motions and hence cause diffraction intensities to decrease at a rate characteristic of a material's phonon–phonon scattering and thermalization time. If materials also exhibit a short carrier lifetime through nonradiative recombination, the bandgap energy per carrier pair is further released to the lattice, which leads to more atomic motions.[26] The lack of notable intensity changes for $Cs_3Bi_2I_9$ in the first few tens of ps, however, signifies both the breakdown of the band picture and a long carrier lifetime. Thus, the increase in the atomic motions, especially the randomized thermal motions, is limited in the first few tens of ps. This finding is in fact fully consistent with what the early-time TR results have revealed about the ultrafast formation of self-trapped excitons and no free carriers, which strengthens the view of $Cs_3Bi_2I_9$ as a "molecular" crystal not only from the structural but also behavioral standpoints. In this regard, a structural modulation of the ionic lattice for the trapped carriers may retain appreciable excess energy, as depicted by the displaced harmonic oscillator diagram with a large Huang–Rhys factor.[18b] Moreover, the long-lasting excitons cause an electronic stress that leads to the formation of CAPs observed in TR.

Second, we further note that the occupation number of the LO phonons involved in the initial coupling and trapping of carriers should be limited in the first 10 ps. Shown in Figure 4b are the simulated diffraction changes as a result of select major optical phonon modes that have atomic displacements along the $c$ axis, with $A_1'$ terminal Bi–I symmetric stretch (in-phase and out-of-phase between the two $[Bi_2I_9]^{3-}$ units), $A_2''$ terminal Bi–I symmetric stretch (in-phase and out-of-phase), $E'$ and $E''$ terminal Bi–I asymmetric stretch, and in-phase $A_1'$ bridge Bi–I symmetric stretch.[18b, 27] A 5% range of the Bi–Bi distance of 4.042 Å is considered for the movement of a single Bi atom, which is ~3/4 of the root-mean-square displacement $\sqrt{B/8\pi^2}$ deduced from the equilibrium thermal factor $B$ at low temperature.[28] To have a negligible effect on the $(00l)$ intensities, the increase in the optical phonon populations and consequently the atomic displacements needs to be small. This appears to in contrast with the extent of the average optical-phonon atomic displacement suggested by the large Huang–Rhys factor of $S = 79.5$ should the lattice fully relaxes:[18b] considering the in-phase $A'_1$ terminal Bi–I symmetric stretch at 146.6 cm$^{-1}$ ($\omega_{LO}$ is the corresponding angular frequency), the carrier–phonon coupling-induced shift $\Delta$ can be estimated by $S = \omega_{LO}\Delta^2/2\hbar$, which gives a sub-Å value that is apparently too large according to the UED results. Thus, this finding provides the structural evidence for a shared and less dominant role played by the LO phonons' initial coupling to the photoinjected carriers. A similar view was also suggested for bismuth double perovskite $Cs_2AgBiBr_6$, although derived from optical spectroscopic data only.[10]

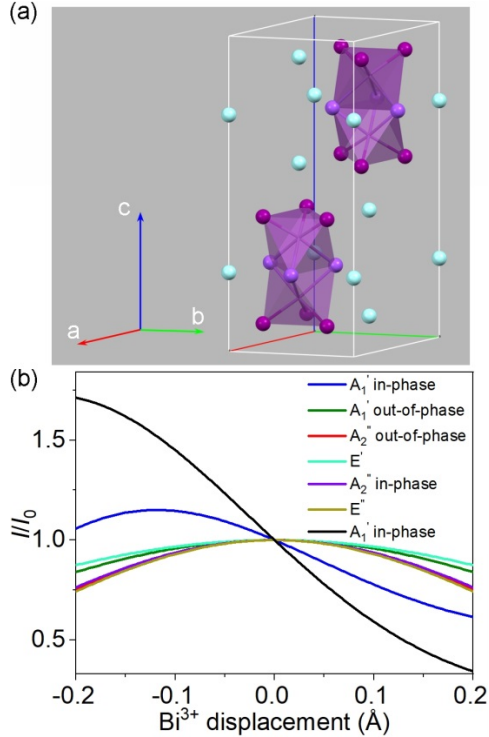

**Figure 4.** Structure of $Cs_3Bi_2I_9$ and diffraction simulations. (a) Unit cell showing the two $[Bi_2I_9]^{3-}$ units with Cs in light blue, terminal I in dark purple, and bridge I in light purple. The bismuth atoms (not shown) are at the centers of the octahedral frameworks. (b) Simulated changes of the (0 0 24) diffraction intensity as a result of lattice distortions following select optical phonons. An atomic displacements of more than a few pm should lead to a noticeable intensity decrease in UED.

Third, the diffraction intensity decrease $\ln[I(t)/I_0]$ at longer times and its linear dependence on $s_\parallel^2$ indicate the increased thermal motions on the sub-ns to ns time. The thermal energy must come from the carriers. Thus, the time constant of ~450 ps in Figure 3b is attributed to the nonradiative recombination time of the trapped excitons at the photoinjection level of ~$10^{20}$ cm$^{-3}$; it is also noted that an exponential-decay component exists on the sub-ns to ns time in low-temperature TR (Figure 2b). A similar lifetime has been reported although at a much lower carrier density.[29] Neglecting carrier diffusion and assuming that most photocarriers undergo the nonradiative process to release their energy to the lattice, we estimate from Figure 3c ~0.038 Å as the increase in the root-mean-square out-of-plane atomic displacement at the sample surface. This value is comparable to the initial electronic lattice strain of $2.3 \times 10^{-3}$ at the surface deduced from the CAP modulations in the TR data, which is ~0.05 Å along $c$ per unit cell (see Supplementary Information).

Thus, a picture emerges for the photodynamics of $Cs_3Bi_2I_9$ after both the TR and UED results are taken into account. The fact that the injected photocarriers are essentially localized from the beginning coincides well with the quasi-0D nature of unconnected $[Bi_2I_9]^{3-}$ units and the electronic wavefunctions involved in the photoexcitation, from mainly iodide $p$ orbitals to bismuth/iodide $p$ molecular orbital states.[11] Such electronic change is accompanied by local structural deformation and/or population of high-frequency optical phonons centered on the excited $[Bi_2I_9]^{3-}$ units (at a density of ~8% near the sample surface) in the first several ps, although the extent should be limited, up to a few pm per atom no more than ~1% of the Bi–I bond length, based on the UED result at early times. Moreover, an electronic stress via the deformation potential coupling is developed during this time up to ~10 ps, whose duration can be understood as the time needed to transition from the molecular-type excitons to lattice distortions coordinating all ions across a few unit cells, hence of the acoustic-phonon type. We note that a similar picosecond time scale was reported for the development of nanoscale polaronic distortions in $CH_3NH_3PbBr_3$.[5] Here, the electronic stress and the exponentially decreasing density profile of the long-lived excitons result in a strain impulse, which is launched in the surface region and propagates into the bulk at the longitudinal speed of sound. This CAP lattice distortion at the surface is about 5 pm per unit cell, which is of a similar magnitude as the earlier molecular deformation. Lastly, the lattice temperature is elevated with more randomized atomic motions, as more and more carriers annihilate and transfer their energy to the lattice via nonradiative recombination at longer times on the sub-ns to ns scale. Incidentally, the Debye–Waller model yields a motion increase of almost the same magnitude as the strain distortion. These results suggest a comparable extent of atomic motions throughout the structural evolution from the initial directional nudges by the local excitations to the directional distortions by the CAPs to finally the incoherent, nondirectional random motions.

**Conclusion**

In summary, our results cross-examining time-resolved optical and direct structural measurements provide a detailed look at the coupling of photoinjected carriers to the crystal structure in bismuth-based halide perovskites. Compared to lead-halide perovskites with a connected network, the photodynamics of $Cs_3Bi_2I_9$ is more of a molecular type due to the quasi-0D nature of the isolated $[Bi_2I_9]^{3-}$ units, which causes the electronic band picture to become inadequate. We find a limited coupling to the optical phonons and structural deformation in the first several ps. However, in the same duration of ~10 ps, the carrier-induced electronic stress via the deformation potential coupling is developed, leading to the launch of a propagating strain field as prominent coherent acoustic phonons. The sub-ns to ns decay of photocarriers eventually releases most of their energy and heats up the lattice to increase random atomic thermal motions. While the stronger self-trapping of carriers in bismuth-based MHPs compared to LHPs may pose a crucial limitation on their potential photovoltaic applications, this dynamic behavior with a long carrier lifetime may make them suitable for other uses such as radiation detection.

**Experimental Section**

Single crystals of $Cs_3Bi_2I_9$ were grown using the inverse temperature crystallization technique, and their characterizations by scanning electron microscopy, energy dispersive x-ray spectroscopy, and x-ray photoelectron spectroscopy have been reported previously.[17b] To ensure the best surface quality for both TR and reflection UED measurements, mm-thick crystals were cleaved to produce a smooth (001) surface, which was quickly rinsed by chlorobenzene prior to the loading of the samples into the vacuum chamber assembly. All measurements were conducted in ultrahigh vacuum with a base pressure of the order of $10^{-10}$ torr to prevent the degradation of the cleaved sample surfaces.

Details about the TR and reflection UED apparatus has been previously described.[25b, 25d, 30] Briefly, the Yb:KGW regeneratively amplified laser system delivers a fundamental output of 170-fs pulses at 1030 nm. The 515-nm (2.41 eV) photoexcitation pulses produced by second harmonic generation (SHG) of the fundamental output were used in both the TR and UED measurements. A low photoinjection repetition rate of no more than 250 Hz was used in order to avoid optical surface damages. For TR, the fundamental output was used as the optical probe beam, whereas the excitation beam was mechanically chopped at half of the probe beam repetition rate. The surface-reflected probe beam was detected by a silicon photodiode connected to a lock-in amplifier for a phase-sensitive detection. For UED, another stage of SHG using a fraction of the 515-nm beam produced the ultraviolet (257 nm) pulses, which were focused on a $LaB_6$ emitter tip to generate photoelectron pulses accelerated to 30 keV. The electron diffraction images of the samples were produced at a grazing incidence angle and captured by an intensified CMOS camera assembly. A beam-front tilt setup was used to optimize the overall system response time.

Full structure optimizations and total energy calculations employed the Vienna *ab initio* simulation package (VASP) within the DFT framework.[31] The plane-wave basis set and projector augmented-wave (PAW) potentials were used to define the electronic wave functions.[32] The crystal structure was first optimized using the Perdew-Burke-Ernzerhof (PBE) exchange and correlation functional[33] with the electronic convergence criteria set to $1\times10^{-8}$ eV while the ionic relaxation was set to $1\times10^{-2}$ eV/Å. A cutoff energy of 500 eV was used and the integration of the first Brillouin zone was carried out using a Monkhorst-Pack *k*-point grid of 9×9×3. Exchange and correlation was additionally described using the screened hybrid functional, HSE06.[34] The HSE06 functional was then implemented to correct for the significant underestimation of the PBE band gap with a Monkhorst-Pack *k*-point grid set to 4×4×2 and the electronic convergence criteria set to $1\times10^{-8}$ eV. Phonon dispersion curves were calculated using the PHONOPY package, which uses the modified Parlinski-Li-Kawazoe *ab initio* force constant method.[35] Spin-orbit coupling was included in all calculations.

**AUTHOR INFORMATION**

**Author Contributions**

D.-S.Y. conceived and supervised the project and wrote the manuscript; X.H. carried out the measurements and analyzed the results; N.K.T. and S.S. provided the single-crystalline samples;

J.B. conducted the electronic band structure and phonon mode calculations. All authors were involved in the discussion of the results and contributed to the manuscript.

**Notes**

The authors declare no competing financial interest.

## ACKNOWLEDGMENTS

This research was primarily supported by the R. A. Welch Foundation (E-1860). X.H. and the instrumental implementation of the pulse-front tilt scheme were partly supported by a National Science Foundation CAREER Award (CHE-1653903).

Supporting Information

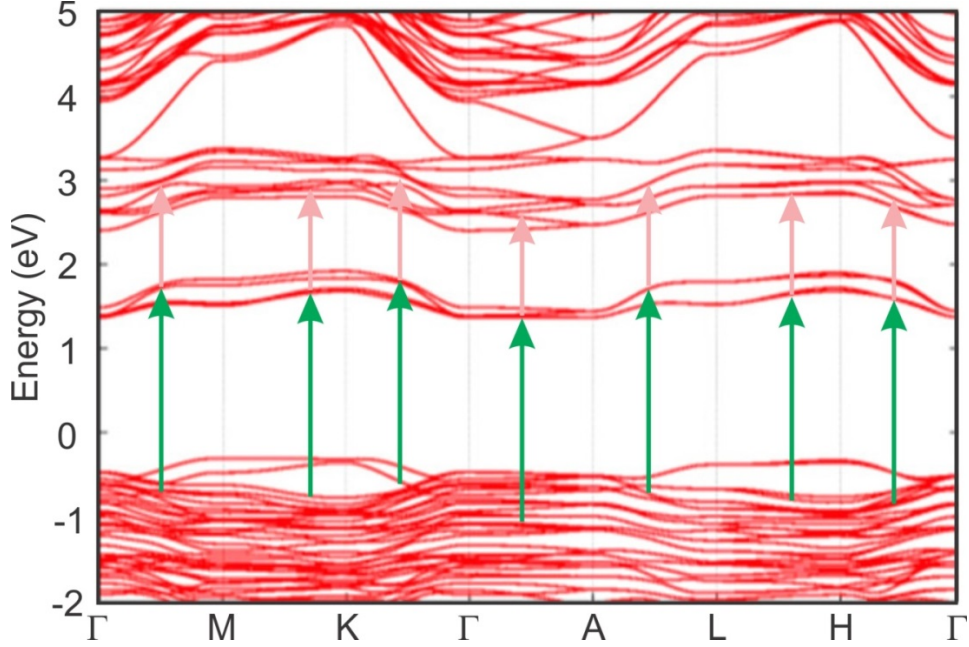

**Figure S1.** Photoexcitation by 515-nm light (green arrows) and probing of the free carriers in the band picture at 1030 nm (pink arrows). The band structure is adapted from Ref. [S1].

Estimate of the anticipated initial transient reflectivity decrease based on free-carrier absorption

At non-absorbed wavelengths, the effective observation depth of a probe beam for transient reflectivity (TR) measurements is given by $d_{\text{pr}} = \lambda_{\text{pr}}/4\pi|n_0|$, where $n_0$ is the refractive index of the material at the probe wavelength $\lambda_{\text{pr}}$.[S2] The above-gap excitation of $Cs_3Bi_2I_9$ at 515 nm leads to the initial photoinjection density profile $N(z)$ as a function of depth $z$

$$N(z) = N(0)e^{-\alpha z} = \frac{F(1-R)\alpha}{h\nu} e^{-\alpha z},$$

where $F$, $R$, and $\alpha$ are the fluence, reflectivity, and absorption coefficient of the excitation beam with the photon energy $h\nu$ = 2.41 eV. With $R \sim 0.145$ calculated from the index of refraction (as an upper bound for the experiments) at 515 nm, an estimate of the average carrier density within the depth $d_{\text{pr}}$ is $N_{\text{avg}} \cong 1.2 \times 10^{20}$ cm$^{-3}$ for $F$ = 0.94 mJ/cm$^2$. The initial refraction index change $\Delta n_{\text{FC}}$ due to free-carrier absorption is given by

$$\Delta n_{\text{FC}} = -\frac{e^2 \lambda_{\text{pr}}^2}{8\pi^2 c^2 \epsilon_0 n_0}\left(\frac{N_{\text{avg}}}{m_e} + \frac{N_{\text{avg}}}{m_h}\right),$$

where $e$ is the charge of a proton, $c$ the speed of light, $\epsilon_0$ the permittivity of free space, and $m_e$ and $m_h$ are the effective masses of electrons and holes, respectively. We note that sizable differences in the $m_e$ and $m_h$ values are seen in the literature.[S1, S3] Furthermore, due to the degenerate carrier concentrations populating the entire three-dimensional density of states, we consider the weighted geometric mean of the parallel ($\parallel$) and perpendicular ($\perp$) values for the effective masses,[S4] i.e.,

$$m_{\mathrm{D}e} = \left(m_{e,\parallel} m_{e,\perp}^2\right)^{1/3} \quad \text{and} \quad m_{\mathrm{D}h} = \left(m_{h,\parallel} m_{h,\perp}^2\right)^{1/3}.$$

The effective reduced mass $\mu_{\mathrm{EX}}$ is then calculated by

$$\mu_{\mathrm{EX}}^{-1} = m_{\mathrm{D}e}^{-1} + m_{\mathrm{D}h}^{-1}.$$

Hence, with the different density-of-state effective masses, the theoretical initial TR change due to the plasma effect is found to be −2.8 to −7.4% if photoinjected electrons and holes were free carriers. However, the Saha equation gives the ratio $n_{e,h}^2/n_{\mathrm{EX}}$ of the order of $10^{13}$ to $10^{14}$ cm$^{-3}$ at room temperature and of $10^4$ to $10^5$ cm$^{-3}$ at 115 K, which means that the fraction of free carriers is <0.1% and 10s ppb, respectively. Therefore, both the experimental results and theoretical considerations lead to the conclusion that ultrafast trapping of photoinjected carriers is at work.

Theoretical calculation of the TR periodic modulations and the strain at the surface

We use Equations 14 and 16 of Ref. [S5] to calculate the oscillatory reflectivity change due to CAPs, where $\tilde{n}_1 = \sqrt{n_0^2 - \sin^2\theta_{\mathrm{in}}}$ is used for the effective refractive index at the probe geometry, the probe light wave number is $k \equiv 2\pi/\lambda_{\mathrm{pr}}$, and the deformation potential $a_{cv}$ (including both the valence- and conduction-band contributions) serves as a fitting parameter. With the different energy-dependent dielectric constants reported in the literature, we obtain $a_{cv}$ to be ~2 eV from the match of the initial amplitude of the periodic modulations (Figure 1c). This value compares reasonably well with the theoretical expression $a_{cv} = -B(\partial E_g/\partial p)$ where the bulk modulus $B \sim 11.7$ GPa is estimated from the experimentally obtained elastic constant $C_\parallel = \rho v_{s,\parallel}^2 = 3 \cdot (1-\sigma)/(1+\sigma) \cdot B$ with the longitudinal speed of sound $v_{s,\parallel} = 1.94$ km/s, the mass density $\rho = 5.02$ g/cm$^3$, and the Poisson's ratio $\sigma \approx 0.3$; $\partial E_g/\partial p$ is estimated to be about −0.3 eV/GPa theoretically or −0.03 eV/GPa experimentally.[S6]

Furthermore, the electronic lattice strain at the surface is[S5]

$$\eta(0) = N(0) a_{cv}/\rho v_{s,\parallel}^2.$$

The resulting longitudinal strain of 2.3×10⁻³ gives a lattice change of ~0.05 Å per unit cell.

Estimate of the coupling strength between carriers and Fröhlich-type optical phonons

The Fröhlich coupling constant can be calculated by

$$\alpha_{e,h} = \frac{e^2}{\hbar} \cdot \frac{1}{4\pi\epsilon_0}\left(\frac{1}{\epsilon_\infty} - \frac{1}{\epsilon_s}\right)\sqrt{\frac{m_{e,h}}{2\hbar\omega_{LO}}},$$

where $\hbar = h/2\pi$ is the reduced Planck constant, $\epsilon_\infty$ and $\epsilon_s$ are the optical and static dielectric constants, and $\omega_{LO}$ is the angular frequency of a characteristic phonon mode. Considering the $A_1'$ symmetric stretching of terminal Bi–I bonds at 146.6 cm⁻¹,[S7] the ranges of $\alpha_e$ and $\alpha_h$ are found to be 1.6–4.5 and 2.4–4.9, respectively, with the use of different reported values of $m_e$, $m_h$, $\epsilon_\infty$, and $\epsilon_s$ (Table S1).

**Table S1.** Parameters of $Cs_3Bi_2I_9$ used in the theoretical estimates.

|  | Value | Ref. |
|---|---|---|
| $\alpha$ | 6.7×10⁴ cm⁻¹ | [S1] |
| $n_0$ | 2.20 | [S8] |
| $v_{s,\parallel}$ | 1.941×10⁵ cm/s | This work |
| $\partial E_g/\partial p$ | eV/GPa |  |
| $\partial\epsilon/\partial E$ | 0.66 eV⁻¹ at $\lambda_{pr}$ | [S9] |
|  | 1.13 + 0.24i eV⁻¹ at $\lambda_{pr}$ | [S10] |
| $a_{cv}$ | 1.78 or 2.33 eV | This work |
| $m_{De} = (m_{e,\parallel}m_{e,\perp}^2)^{1/3}$ | 1.512 $m_0$ | [S3a] |
|  | 1.026 $m_0$ | [S3b] |
|  | 0.446 $m_0$ | [S1] |
| $m_{Dh} = (m_{h,\parallel}m_{h,\perp}^2)^{1/3}$ | 1.759 $m_0$ | [S3a] |
|  | 1.084 $m_0$ | [S3b] |
|  | 1.002 $m_0$ | [S1] |
| $\bar{\epsilon}_\infty = (\epsilon_{\infty,\parallel}\epsilon_{\infty,\perp}^2)^{1/3}$ | 4.10 | [S3a] |
|  | 5.03 | [S3a] |
| $\bar{\epsilon}_s = (\epsilon_{s,\parallel}\epsilon_{s,\perp}^2)^{1/3}$ | 9.1 | [S3a] |
|  | 9.16 | [S11] |

Changes of the (00*l*) diffraction intensities due to intracell atomic displacements

The coupling of self-trapped excitons to optical phonons may lead to atomic displacements within a unit cell, which in turn may cause the intensity of a Bragg diffraction to change. We consider the kinematic scattering theory to simulate the possible changes,

$$I \propto \left|\sum_j f_j^{(e)}(s) \exp(-2\pi i \vec{s} \cdot \vec{r}_j)\right|^2 = \left|\sum_j f_j^{(e)}(lc^*) \exp\left(-2\pi i lc^* \cdot (z + \Delta z_j)\right)\right|^2,$$

where $f_j^{(e)}(s)$ is the atomic scattering factor of the *j*-th atom as a function of $s$, $\vec{s}$ is the momentum transfer vector whose magnitude is $s = lc^*$ for the current study, and $\vec{r}_j = (x_j + \Delta x_j, y_j + \Delta y_j, z_j + \Delta z_j)$ is the *j*-th atom's position with the transient movement within a unit cell. Thus, the atomic displacements along the *c* axis (i.e., the surface normal direction) given by different optical phonon modes are used. By symmetry, the movements of the four $Bi^{3+}$ ions are of the same magnitude but may have the same or different signs. Hence, a single $Bi^{3+}$ displacement value can be used to represent the linearly-scaled movements of all 28 atoms in a unit cell. We note that different modes cause different extents of the intensity change, as small as close to 0 for the modes with mostly in-plane motions to as large as more than 30% for the modes with sub-Å vertical movements. Based on the nature of the photoexcitation, we focus on the important modes associated with more prominent $[Bi_2I_9]^{3-}$ intra-unit motions.

As shown in Figure 4, most optical phonon modes lead to a symmetric intensity decrease profile for movements in both the positive and negative directions. This means the anticipation of a diffraction intensity decrease from an averaged result considering a full vibration cycle, since the UED measurements did not have the temporal resolution to resolve the instantaneous change. Hence, the experimentally observed little decrease in the diffraction intensity signifies that the displacements cannot be too large beyond a few pm. In contrast, the modes of in-phase $A_1'$ terminal Bi–I symmetric stretch and in-phase $A_1'$ bridge Bi–I symmetric stretch exhibit asymmetric intensity change profiles (Figure 4, dark blue and black curves). An average of the intensity change over a vibration period would yield basically the same diffraction intensity as long as still within the approximately linear region near the center, i.e. zero displacement.[S12] Thus, strictly speaking, it is possible to have more population of these two optical phonons. Such a result appears to be consistent with the nature of a local excitation in a $[Bi_2I_9]^{3-}$ unit.

Estimate of the Debye temperature from the ultrafast electron diffraction data

Considering the Debye–Waller effect for the increased noncoherent motions of ions, the time-dependent intensity change

$$\ln\frac{I_0}{I(t)} = 2W(t) = \Delta\langle(\vec{q}\cdot\vec{u}(t))^2\rangle = 4\pi^2 s_\parallel^2 \cdot \Delta\langle u_\parallel^2(t)\rangle$$

where $I_0$ and $I(t)$ are, respectively, the diffraction intensities of a diffraction spot before the zero of time and at time $t$ after excitation, $W$ the Debye−Waller factor, $\vec{u}$ the atomic displacement vector, $\vec{q} = 2\pi\vec{s}_\parallel$ the scattering vector of the diffraction, and $s_\parallel = l\cdot c^*$ the momentum transfer parallel to the surface normal with $c^* = 1/c = 0.0472$ Å$^{-1}$ being the corresponding reciprocal primitive cell length. We estimate an upper bound for the Debye temperature $\Theta_D \cong 228$ K from the out-of-plane data using the slope of the linear fit in Figure 3c and the equation

$$4\pi^2\Delta\langle u_\parallel^2(T)\rangle = \frac{3h^2}{\bar{m}k_B\Theta_D}\left[\left(\frac{T}{\Theta_D}\right)^2\int_0^{\frac{\Theta_D}{T}}\frac{xdx}{e^x-1} - \left(\frac{T_0}{\Theta_D}\right)^2\int_0^{\frac{\Theta_D}{T_0}}\frac{xdx}{e^x-1}\right]$$

where $u_\parallel(T)$ is the atomic displacement along the out-of-plane axis, $\bar{m}$ the average atomic mass, $k_B$ the Boltzmann constant, and $T_0 = 115$ K is the experimental base temperature. The elevated temperature $T = 193.6$ K, an upper bound, is estimated based on the assumption that all the absorbed photon energy is transferred to the lattice after 2 ns, with the equation

$$T - T_0 = F(1-R)\alpha/\rho C$$

where $C = 0.136$ J/(g·K) is the specific heat at $T_0$ using the calorimetric data of two similar compounds.[S13] The Debye temperature may seem somewhat large for a softer perovskite with a larger unit cell.[S14] However, given the highly anisotropic properties of Cs$_3$Bi$_2$I$_9$, if the out-of-plane longitudinal speed of sound (instead of the average value) is used, an estimate of $\Theta_D \cong 161$ to 171 K is obtained according to the model[S15]

$$\Theta_D = \frac{h}{k_B}\left(\frac{3N}{4\pi V}\right)^{1/3} v_s$$

where $N$ and $V$ are the number of atoms in and the volume of a unit cell, respectively. We note that these Debye temperature estimates are reasonably close.